\documentclass[aps,prb,twocolumn,groupedaddress,showpacs,superscriptaddress,amssymb,amsmath]{revtex4-1}
\usepackage[utf8]{inputenc}
\usepackage{graphicx}
\usepackage{tabularx}
\usepackage{xcolor}
\usepackage{amsmath}

\usepackage{comment}

\usepackage{dcolumn}
\usepackage{hyperref}
\hypersetup{
    colorlinks=true,
    linkcolor=blue,
	citecolor=cyan
}
\usepackage{bm}
\usepackage{epsf}
\usepackage{subfigure}
\usepackage{braket}
\usepackage{tensor}
\usepackage{soul}

\usepackage{mathrsfs}
\usepackage{mathtools}
\usepackage{multirow}

\usepackage{amsfonts}

\newcommand{\be}{\begin{equation}}
\newcommand{\ee}{\end{equation}}
\newcommand{\bea}{\begin{eqnarray}}
\newcommand{\eea}{\end{eqnarray}}

\makeatletter
\newcommand*\rel@kern[1]{\kern#1\dimexpr\macc@kerna}
\newcommand*\widebar[1]{%
  \begingroup
  \def\mathaccent##1##2{%
    \rel@kern{0.8}%
    \overline{\rel@kern{-0.8}\macc@nucleus\rel@kern{0.2}}%
    \rel@kern{-0.2}%
  }%
  \macc@depth\@ne
  \let\math@bgroup\@empty \let\math@egroup\macc@set@skewchar
  \mathsurround\z@ \frozen@everymath{\mathgroup\macc@group\relax}%
  \macc@set@skewchar\relax
  \let\mathaccentV\macc@nested@a
  \macc@nested@a\relax111{#1}%
  \endgroup
}
\makeatother

\begin{document}
\newcolumntype{M}[1]{>{\centering\arraybackslash}m{#1}}
\title{Anomalous transport in long-ranged open quantum systems}
\author{Abhinav Dhawan}
\email{abhinav.dhawan@students.iiserpune.ac.in}
\affiliation{Department of Physics,
		Indian Institute of Science Education and Research, Pune 411008, India}
\author{Katha Ganguly}
\email{katha.ganguly@students.iiserpune.ac.in}
\affiliation{Department of Physics,
		Indian Institute of Science Education and Research, Pune 411008, India}


\author{Manas Kulkarni}
  \email{manas.kulkarni@icts.res.in}
\affiliation{International Centre for Theoretical Sciences, Tata Institute of Fundamental Research,
Bangalore 560089, India}

\author {Bijay Kumar Agarwalla}
\email{bijay@iiserpune.ac.in}
\affiliation{Department of Physics,
		Indian Institute of Science Education and Research, Pune 411008, India}
		\email{bijay@iiserpune.ac.in}

	\date{\today}

\begin{abstract}
We consider a one-dimensional fermionic lattice system with long-ranged power-law decaying hopping with exponent $\alpha$. The system is further subjected to dephasing noise in the bulk. We investigate two variants of the problem: (i) an open quantum system where the setup is further subjected to boundary reservoirs enabling the scenario of a non-equilibrium steady state charge transport, and (ii) time dynamics of an initially localized single particle excitation in the absence of boundary reservoirs. In both variants, anomalous super-diffusive behavior is observed for $1< \alpha< 1.5$, and for $\alpha> 1.5$ the setup is effectively short-ranged and exhibits conventional diffusive transport. Our findings are supported by analytical calculations based on the multiple scale analysis technique that leads to the emergence of a fractional diffusion equation for the  density profile. Our study unravels an interesting interplay between long-range interaction and dephasing mechanism that could result in the emergence of unconventional behaviour in open quantum systems.  
\end{abstract}

\maketitle 

{\it Introduction.} Quantum transport in low dimensional systems has been an active area of research specially due to the evidence of unconventional or anomalous behaviour \cite{abhishek2008, RevModPhys.93.025003,  RevModPhys.94.045006, purkayastha2018anomalous,subdiffusive_long, subdiffusive_transfer,KPZ2019,PhysRevLett.108.180601,PhysRevLett.121.230602,Bar_Lev_2017,PhysRevX.13.011033,PhysRevB.86.125118,PhysRevLett.114.160401,rojo2024anomalous,chen2023superdiffusive}. Research in this direction is not only of fundamental importance but can potentially have technological applications \cite{Jordens2008,Greiner2002,Jepsen2020,Atala2014,Syassen2008,Sieberer_2016,PhysRevLett.88.094302,PhysRevLett.93.184301}. Remarkable experimental progress \cite{R_Blatt2022,Bonato2016,PhysRevLett.77.4728,Schafer2020,bloch2012quantum,PRL10qubit2019,Barredo2016,blatt2012quantum,britton2012engineered} in quantum devices and architecture has made it possible to realize one-dimensional systems which can in principle exhibit conductance ${\cal G}$ that scales differently from the conventional system size scaling ${\cal G} \propto 1/N$ with $N$ being the system size. Such a departure from diffusive behaviour is termed anomalous and conductance in such cases scales as ${\cal G} \propto 1/N^{\delta}$ where $0 < \delta <1$ corresponds to super-diffusive transport and $\delta >1$ corresponds to sub-diffusive transport. Another widely employed alternate approach to classify transport is by studying the spread of wave-packets in the system \cite{purkayastha2018anomalous, schuckert2020nonlocal, PhysRevLett.119.046601,prelovvsek2018transient,BarLevAnderson,Bhakuni_noise}.  
The exponent associated with the scaling collapse of the wave function spread dictates the nature of anomalous transport or lack thereof. Often, there can be interesting relations between this exponent and the one associated with conductance scaling with system size \cite{Abhishek_Levy, PhysRevLett.91.044301, PhysRevLett.112.040601}. 

Such anomalous transport is often predicted in setups with correlated or uncorrelated disorder \cite{PhysRevLett.124.130604,PhysRevB.106.045417,PhysRevB.100.115311,PhysRevB.88.014201,PhysRevResearch.3.033257,poon2023anomalous}. It has also been reported in interacting many-body quantum systems \cite{PhysRevB.102.115121,PhysRevB.99.241113,brighi2024anomalous,PhysRevLett.127.230602,PhysRevX.13.011033,PhysRevLett.114.100601} or quantum systems subject to environmental effects \cite{PhysRevLett.119.046601,prelovvsek2018transient,BarLevAnderson,Bhakuni_noise}. Despite a plethora of progress, a rigorous microscopic understanding of anomalous transport is still lacking. This is primarily due to the lack of simple platforms where both analytical calculations and state-of-the-art numerical simulations can be performed. Moreover, it is worth emphasising that most works studying anomalous transport consider short-range interacting systems. Some progress has recently been made in long-range systems where unconventional behaviour is reported \cite{subhajit_long_range,iubini2022hydrodynamics}. Therefore, it is of paramount importance to explore the hidden mechanisms in setups with long-range coherent coupling, which results in anomalous behaviour. One such avenue is when long-range systems are subjected to dephasing mechanism and this is the central platform used in our work.  

In this work, we study quantum transport properties in a one-dimensional long-range fermionic lattice system with the long-range hopping exponent denoted by $\alpha$. This setup is further sujected to dephasing noise, often called as B\"uttiker voltage probes (BVP), that acts at each lattice site. We first study the non-equilibrium steady state (NESS) scenario by connecting the setup by two boundary fermionic reservoirs one at each end. These boundary reservoirs result in the establishment of a NESS current. The given platform is amenable to exact results (non-perturbative and non-Markovian) in NESS. We investigate the system size scaling of conductance and observe super-diffusive transport regime when $1<\alpha<1.5$ and conventional diffusive regime when $\alpha>1.5$. We provide a compelling evidence for a relationship between system size scaling exponent of conductance ($\delta$) and long-range hopping exponent ($\alpha$). 
We then characterize the transport by performing quantum dynamics study of single-particle excitation in the absence of the fermionic boundary reservoirs but retaining the dephasing mechanism. The exponent ($\eta$) associated with the space-time collapse of the single particle density profile is found to be super-diffusive for $1<\alpha<1.5$ and diffusive for $\alpha>1.5$. We support these findings by obtaining a fractional diffusion equation for the density profile following a multiple scale analysis technique \cite{Bender1999}. We further obtain a relationship between $\eta$ and $\alpha$. This along with the relationship between $\eta$ and $\delta$, known in the context of L\'evy flights \cite{Abhishek_Levy, Iubini_2022}, establishes a relation between $\delta$ and $\alpha$ which is fully in agreement with our NESS analysis. 

\begin{figure}
\includegraphics[width=\columnwidth]{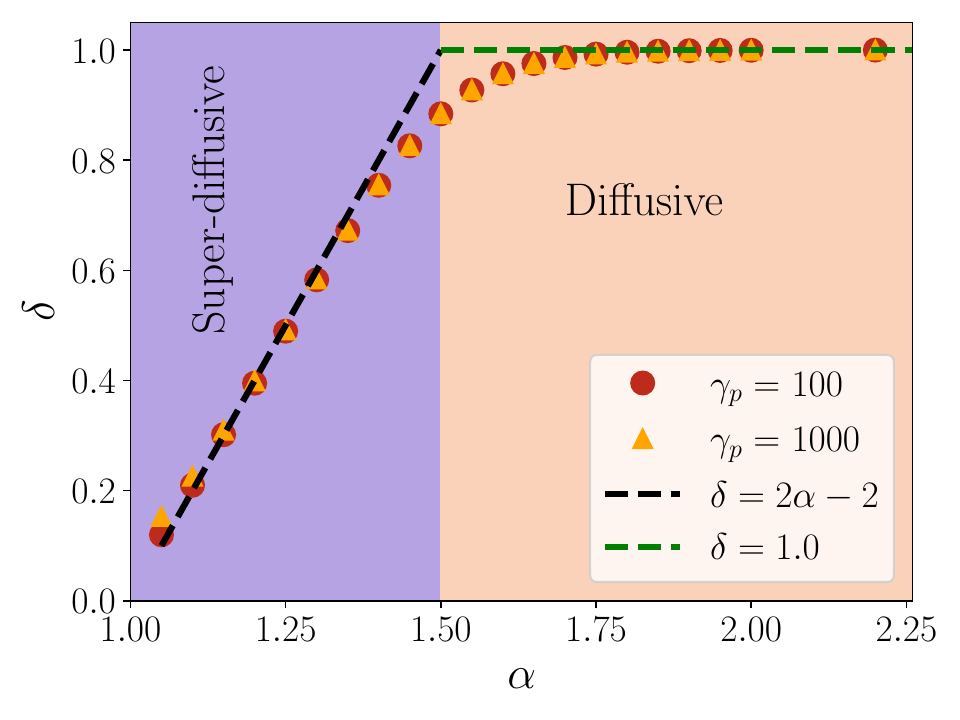}
\caption{Plot for the system size scaling exponent $\delta$ for NESS conductance with long-range hopping exponent $\alpha$. The exponent $\delta$ is extracted by computing conductance for three large system sizes, $N=8192$, $11585$, and $N=16384$. The transition from super-diffusive to diffusive transport regime is observed when $\alpha \sim 1.5$. We plot results for two different choices for $\gamma_p$, $\gamma_p=100, 1000$. We set $\gamma=J=1$. For the plot we choose, $\mu=\epsilon^\alpha_{lb}+0.5$, where $\epsilon^\alpha_{lb}$ denotes the energy corresponding to the lower band edge for the given $\alpha$. However note that, in the strong $\gamma_p$ limit, the scaling exponent $\delta$ is almost independent of the value of the equilibrium chemical potential $\mu$.}
\label{Fig:1}
\end{figure}

{\it Long-range lattice setup and NESS transport.}
We consider a one-dimensional fermionic lattice chain with long-range hopping. The Hamiltonian is given by,  \cite{subdiffusive_long, subhajit_long_range}
\begin{equation}
\label{long-range-Ham}
\hat{{\cal H}}_{S}=-\sum\limits_{m=1}^N \frac{J}{m^{\alpha}} \Bigg[\sum\limits_{r=1}^{N-m}  \hat{c}^{\dagger}_{r}  \hat{c}_{r+m} + \hat{c}^{\dagger}_{r+m}  \hat{c}_{r} \Bigg],
\end{equation}
where $N$ is the system size, $\hat{c}_r^{\dagger} (\hat{c}_r)$ is the fermionic creation (annihilation) operator for the $r$-th site, and $\alpha$ is the long-range hopping exponent. To understand the steady-state transport properties, the lattice chain is further connected to a source and a drain reservoir at its two ends, and these reservoirs are maintained at chemical potentials $\mu_S$ and $\mu_D$, respectively. The interaction between the system and each reservoir, whose strength is denoted by $\gamma$, is chosen to be bilinear and number-conserving. Here, for simplicity, we consider the wide-band limit of the reservoirs which implies frequency independent density of states. In addition to the boundary reservoirs, at each lattice site, we attach B\"uttiker voltage probes (BVP) with uniform coupling strength denoted by $\gamma_p$.  This is done to mimic processes where the phase coherence of particles built during Hamiltonian evolution is lost due to inevitable surroundings \cite{PhysRevB.41.7411,10.1063/1.4944470,PhysRevB.105.134203}. Such an approach is widely employed to understand effective many-body transport \cite{PhysRevB.75.195110,Madhumita2022PRB}. It is to be borne in mind that BVP's are essentially themselves similar to the boundary reservoirs discussed above with the difference being their chemical potentials (denoted by $\mu_i, i=1, 2, \cdots N$ with $i$ stands for the index for the lattice site) are carefully engineered to ensure zero particle NESS current. The temperature however for the boundary reservoirs and the BVP's are always considered to be the same. Note that we do not assume any restriction on the magnitude of the coupling between the system and the boundary reservoirs/probes. We are interested in studying the NESS electronic conductance to characterize transport. We focus here in the linear response regime and set for boundary reservoirs $\mu_D= \mu$, $\mu_S= \mu + \delta \mu$, and for the BVP's $\mu_i = \mu + \delta \mu_i, i=1, 2, \cdots N$. At  zero temperature, the conductance corresponding to the left to right charge current can be exactly obtained as \cite{Madhumita2022PRB}
\begin{eqnarray}
{\cal G}(\mu) &=& \gamma^2 |G_{1N}(\mu)|^2 \nonumber \\
&+& \gamma^2 \gamma_p \sum_{n,j=1}^{N} |G_{Nn}(\mu)|^2 \, {\cal W}_{nj}^{-1}(\mu) \, |G_{j1}(\mu)|^2,
\label{conductance}
\end{eqnarray}
where $G_{ij}(\omega)$ is the matrix element of the $N \times N$ dressed retarded Green's function matrix ${G}(\omega)$ for the lattice, which can be written as,
\begin{equation}
    {G}(\omega) = \Big[\omega \,{I} -{h}_S - { {\bf \Sigma}}_L(\omega) - { {\bf \Sigma}}_R(\omega) -  {{\bf \Sigma}}_{P}(\omega)\Big]^{-1}
\end{equation}
with ${I}$ being the $N \times N$ identity matrix, ${h}_S$ being the single particle $N \times N$ matrix corresponding to the lattice Hamiltonian and defined as $\hat{\cal H}_S = \sum_{l,m=1}^{N} [{h}_S]_{lm} \hat{c}_l^{\dagger} c_m$ in Eq.~\eqref{long-range-Ham},  ${\bf \Sigma}_{\alpha}(\omega), \alpha=L,R,P$ are the self-energy matrices for the left reservoir, right reservoir and the BVPs. The self-energy matrices are diagonal with ${\bf \Sigma}_L(\omega)|_{11}= - i \gamma/2$, ${\bf \Sigma}_R(\omega)|_{NN}= - i \gamma/2$, and ${\bf \Sigma}_P(\omega)|_{jj}= - i \gamma_p/2, \, j=1,2, \cdots N$. The matrix elements of $N \times N$ matrix ${\cal W}(\omega)$ in Eq.~\eqref{conductance} are given by \cite{Madhumita2022PRB} (we suppress the argument in  ${\cal W}$ and $G$ for the sake of brevity) 
\begin{eqnarray}
{\cal W}_{ij} &=& -\gamma_p |G_{ij}|^2 \quad \forall  \quad  n \neq j \nonumber \\
{\cal W}_{ii} &=& \gamma \Big(|G_{i1}|^2 + |G_{iN}|^2 \Big)  + \gamma_p \sum_{j \neq n}^{N} |G_{nj}|^2 \,\, \forall  \,\, n \neq j. \nonumber \\
\label{w-matrix}
\end{eqnarray}
The zero particle NESS current from each of the BVP ensures a unique chemical potential value at each lattice site and is given by \cite{Madhumita2022PRB} 
\begin{equation}
\mu_i = \mu_S + \gamma \, \gamma_p \,(\mu_D- \mu_S) \, \sum_{j=1}^{N} {\cal W}_{ij}^{-1} \, |G_{jN}|^2,
\label{chem-potential}
\end{equation}
where $i=1, 2, \cdots N$ and ${\cal W}$ is given in Eq.~\eqref{w-matrix}. Having the expressions for conductance [Eq.~\eqref{conductance}] and chemical potential [Eq.~\eqref{chem-potential}] in hand, we now provide the results based on extensive numerical simulations. 


In the context of long-range systems, a natural question one may ask is the behavior of system size scaling exponent $\delta$ with respect to long-range hopping exponent $\alpha$. This is demonstrated in Fig.~\ref{Fig:1}. In the limit of large probe coupling strength ($\gamma_p \gg J, \gamma$) we observe a super-diffusive transport regime $(0 < \delta <1)$ for $1< \alpha < 1.5$. Our findings indicate that the relationship between $\delta$ and $\alpha$ in the super-diffusive regime is $\delta \approx 2\alpha -2 $.  We further find a diffusive transport regime i.e., $\delta =1$ for $\alpha > 1.5$. This remarkable crossover in the nature of transport is rooted in the effectively short-ranged interaction when $\alpha>1.5$. In contrast, an interesting interplay between the dephasing noise introduced by the BVPs and the effectively long-ranged hopping ($\alpha<1.5$) gives rise to a faster than diffusive or super-diffusive transport regime. Remarkably similar observations for transport were recently reported for Lindbladian systems \cite{subhajit_long_range}.


\begin{figure}
\includegraphics[width=\columnwidth]{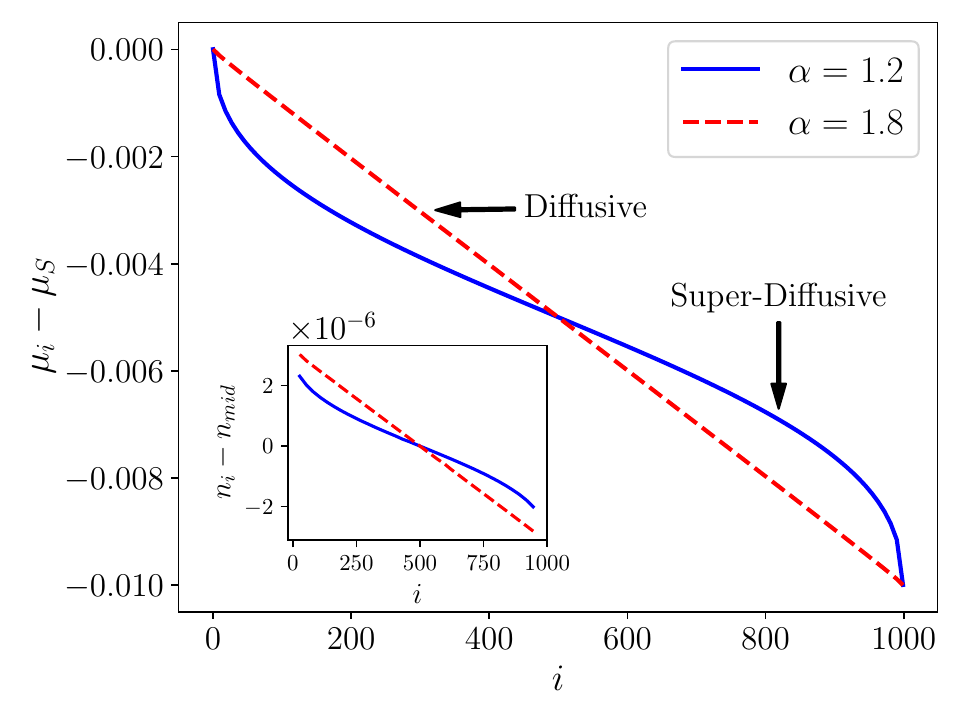}
\caption{Plot for the local chemical potential $\mu_i$ (subtracted by the chemical potential of the source end $\mu_S$) profile as given in Eq.~\eqref{chem-potential} for $\alpha=1.2$ (blue-solid), and $\alpha=1.8$ (red-dashed). The distinct line shapes of the profiles in the two transport regimes can be clearly seen here. We choose the value of $\mu_S$ to be the energy corresponding to the middle of the band, i.e., for $\alpha=1.2$, we choose $\mu_S=-4.87$ and for $\alpha=1.8$, we choose $\mu_S=-1.08$, and $\delta \mu=0.01$. The system size $N=1000$. We choose $\gamma_p=1000$. The inset represents the local occupation number $n_i$ (subtracted by the density at the middle of the lattice $n_{mid}$ to clearly demonstrate the density profile) vs $i$ in the two regimes. $n_{mid}$ depends on the chemical potential and the density of states, which is different in the two regimes.}
\label{fig:2}
\end{figure}


In Fig.~\ref{fig:2}, we demonstrate the local chemical potential profile for two different values of $\alpha$, one within the super-diffusive regime and one within the diffusive regime. For $\alpha>1.5$, we notice a linear shape, which is a hallmark of conventional diffusive transport \cite{Abhishek_Levy}, whereas for $\alpha<1.5$, the shape is nonlinear, which is a fingerprint of anomalous transport \cite{Abhishek_Levy}.  The inset of Fig.~\ref{fig:2} demonstrates the local occupation number in the two distinct regimes. Given the detailed understanding of transport regimes when boundary reservoirs are attached, it is natural to explore the possible relation with the density profile evolution in the absence of reservoirs (but retaining the dephasing mechanism). 

\begin{figure}
\centering
\includegraphics[width=0.49\columnwidth] {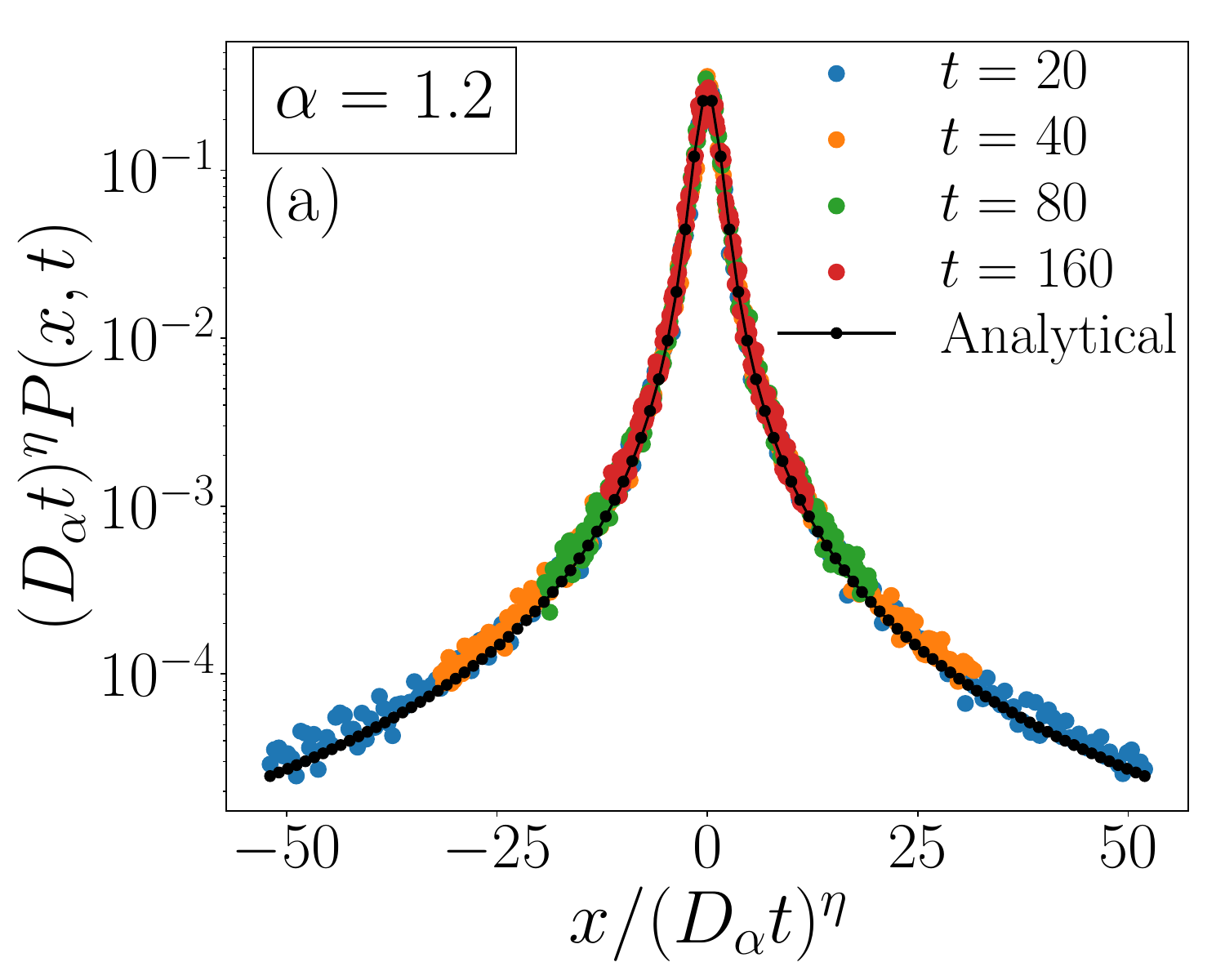}
   \label{fig:1a}
\includegraphics[width=0.49\columnwidth]{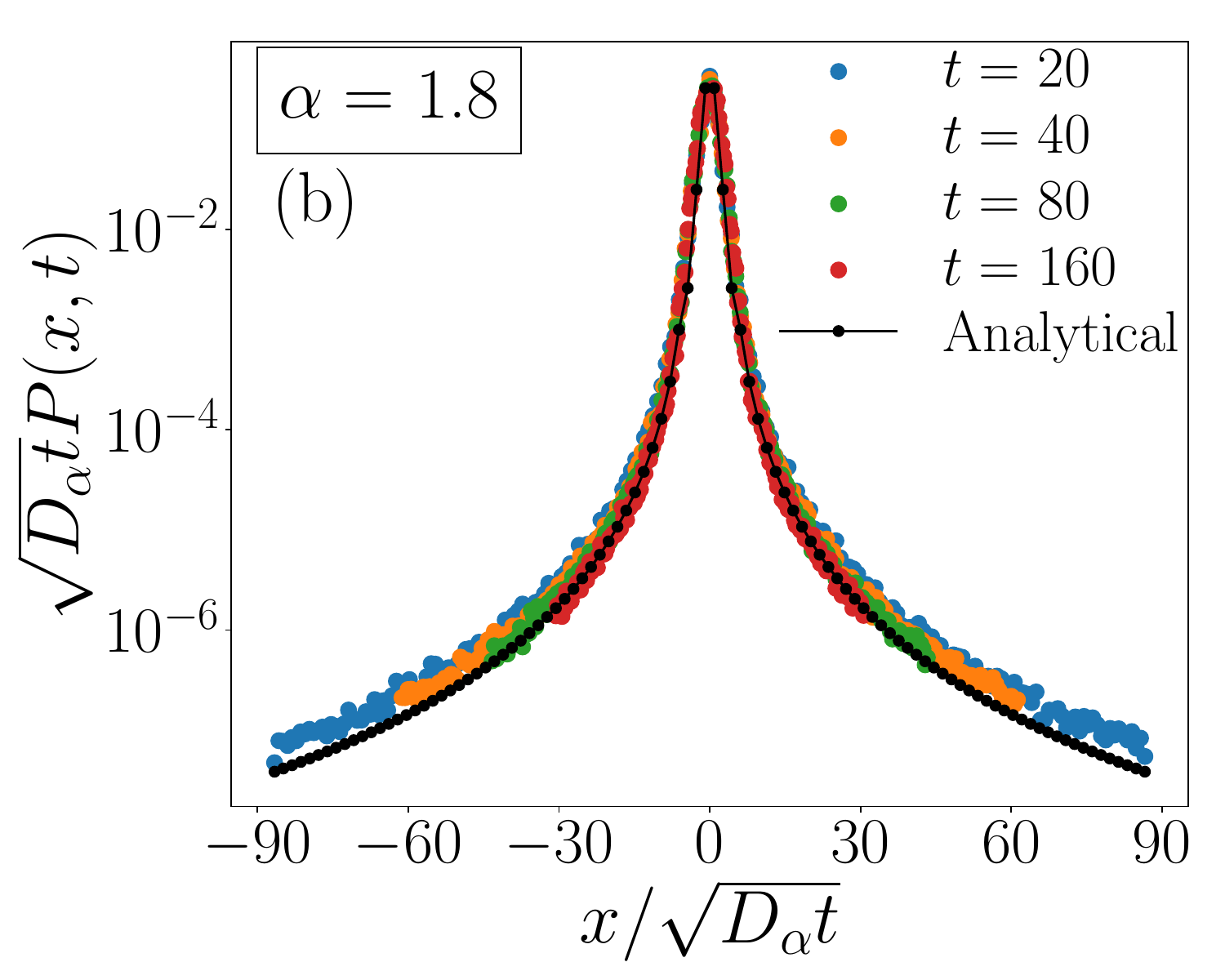}
   \label{fig:1a}
   \caption{(Color online) Time-dynamics shown as a scaling collapse for the single-particle density profile that is initially localized at the middle site of the lattice for (a) $\alpha=1.2$ and (b) $\alpha=1.8$. For (a) the collapse happens for $\eta \approx 1/(2 \alpha -1)$ and for $\alpha=1.2$, $\eta \approx 0.71$ . For both the figures, we choose $N=201, \gamma_p=50$, and noise average over 60 different realizations are performed with $dt=0.01$ (Note, we require $dt\ll {\rm min}\{J^{-1},\kappa^{-1}\}$). The black curves are analytically derived scaling functions as mentioned in Eq.~\eqref{scaling-form} along with finite time corrections \cite{schuckert2020nonlocal} that are crucial to describe the heavy tails in the diffusive regime as shown in (b) which survive for algebraically long times. The corrections explicitly depend on time, and in particular, for (b), the black curve corresponds with $t=160$.}
   \label{fig:collapse1pt2}
\end{figure}

{\it Time dynamics of single particle density profile:} We now study the quantum dynamics of single particle excitation for the long-range lattice setup in the absence of the boundary reservoirs while keeping the dephasing mechanism intact. We model the lattice and this dephasing mechanism by a Lindblad quantum master equation (LQME) \cite{Turkeshi_PhysRevB,gopalakrishnan2017noise,znidaric_2010,subhajit_long_range,non-Gaussian-Demler,tupkary2022fundamental, nathan2020universal,maniscalco2004lindblad,manzano2020short,Thermo-Dario,lindblad1976generators}, given as (setting $\hbar =1$)
\begin{equation}
\frac{d \hat{\rho}}{dt}= -i [\hat{{\cal H}}_S, \hat{\rho}] + \kappa \sum_{i=1}^{N} \Bigg[\hat{n}_i \, \hat{\rho}\, \hat{n}_i - \frac{1}{2}\Big\{\hat{n}_i^2, \hat{\rho}\Big\}\Bigg]
\label{QME}
\end{equation}
where the first term in Eq.~\eqref{QME} is responsible for unitary evolution with $\hat{\cal H}_S$ given by Eq.~\eqref{long-range-Ham}, and the second term mimics the dephasing mechanism with  $\hat{n}_i= \hat{c}_i^{\dagger} \hat{c}_i$ being the number operator corresponding to the $i$-th site. $\kappa$ represents the effective coupling strength characterising dephasing \cite{Turkeshi_PhysRevB,gopalakrishnan2017noise,znidaric_2010,subhajit_long_range,non-Gaussian-Demler,Bhakuni_noise}. Note that, although the lattice Hamiltonian $\hat{\cal H}_S$ in Eq.~\eqref{QME} is quadratic, the open quantum system version as written in Eq.~\eqref{QME} is far from being trivial. This is due to the presence of dephasing terms, which results in the appearance of a quartic term in the LQME \cite{non-Gaussian-Demler}. For such a setup, we are interested in studying the time dynamics of a single-particle density profile $P(x,t)$
which is initially localized at the middle site of the lattice. Note that, here we introduced a new variable $x=i- (N+1)/2$ so as to have the excitation at $x=0$ and we assume the lattice size $N$ to be odd here, without any loss of generality.
Obtaining $P(x,t)$ directly following the LQME in Eq.~\eqref{QME} is computationally expensive since one has to deal with large $N^2 \times N^2$ matrices. This issue can be circumvented by following a unitary unraveling procedure \cite{WISEMAN200191,DSalgado_2002} of Eq.~\eqref{QME} which makes it more computationally feasible. The unraveling is carried out by introducing classical delta-correlated Gaussian noise at each lattice site \cite{PhysRevB.102.100301,Bhakuni_noise}. The total Hamiltonian can therefore be written as 
\begin{equation}
\hat{{\cal H}}(t) = {\cal H}_S + \sum_{l=1}^{N} \xi_l(t)\, n_l, 
\label{noise-H}
\end{equation}
with $\langle \xi_l(t) \rangle = 0$, and $\langle \xi_l(t) \, \xi_p(t') \rangle = \kappa \, \delta_{lp} \, \delta(t-t')$, where recall that $\kappa$ is the dephasing strength. For each noise realization, we perform the dynamics governed by $\hat{\cal H}(t)$ in Eq.~\eqref{noise-H}, and the quantity of interest is obtained by averaging over different noise realizations. For a single noise realization, the single particle density profile is obtained as $P_{\xi}(x,t) = |\psi^{\xi}_x(t)|^2$, with 
\begin{equation}
\psi^{\xi}_{x}(t) ={G}_{\xi}\Big(\frac{N+1}{2}+x,t\Big{|} \frac{N+1}{2},0\Big), 
\label{noise-avg}
\end{equation}
where 
\begin{equation}
G_{\xi}\Big(x_1,t_1 \big{|}x_0, t_0\Big) = \Big\langle x_1 \big{|} {\cal T} e^{-i \int_{t_0}^{t_1} h(t') dt'} \big{|}x_0 \big \rangle
\end{equation}
is the single-particle unitary propagator with $h(t)$ being the single particle Hamiltonian, defined as $\hat{\cal H}(t) = \sum_{l,m=1}^{N} h_{lm}(t) \hat{c}_{l}^{\dagger} \hat{c}_m$, following  Eq.~\eqref{noise-H}. Here ${\cal T}$ is the time-ordered operator. Note that, the variable $x$ goes from $(-N+1)/2$ to $(N\!-\!1)/2$. We evaluate the full propagator by performing infinitesimal time propagation in steps of $dt$ and write 
${\cal T} e^{-i \int_{t_i}^{t_i+dt}  h(t') dt'} \approx  e^{- i \big(h_S dt + \sqrt{dt}\, \mathcal{M} (t_i)\big)}$ where $\mathcal{M}(t_i)$ is a $N\times N$ diagonal noise matrix at time $t_i$ with $m$-th entry of the matrix corresponds to the value of the uncorrelated noise at the $m$th site i.e. $\langle\mathcal{M}_{m,m}(t_i)\rangle=0$ and $\langle\mathcal{M}_{m,m}(t_i)\mathcal{M}_{n,n}(t_j)\rangle=\kappa\, \delta_{mn}\,\delta_{ij}$. Finally, the single-particle density profile at a particular time instant can be obtained by averaging over different noise trajectories, i.e., $P(x,t) = \overline{P_{\xi}(x, t)}$. We first present the numerical results for $P(x,t)$. In Fig.~\ref{fig:collapse1pt2}(a) and (b), we show that the numerics obeys the space-time scaling collapse of $P(x,t)$ of the form 
 \begin{equation}
     P(x,t)= \frac{1}{(D_\alpha t)^{\eta}} \, f\bigg(\frac{x}{(D_\alpha t)^{\eta}}\bigg).
     \label{collapse}
 \end{equation}
In the regime $1 < \alpha < 1.5$, we find  $\eta \approx 1/(2 \alpha -1)$ and for $\alpha>1.5$, we get $\eta \approx 1/2$. In what follows we show that 
one can analytically obtain a fractional diffusion equation for the single particle density profile that provides both the scaling exponents and also the scaling function $f(x/(D_{\alpha} t)^{\eta})$ given in Eq.~\eqref{collapse}. 

To arrive at the fractional-diffusion equation, we use the multiple-scale analysis technique. 
We focus on the correlator $D_{m,n}(t)\equiv \langle \hat{c}_m^\dagger(t)\hat{c}_n(t) \rangle$, where both $m$ and $n$ goes from $(-N+1)/2$ to $(N\!-\!1)/2$.
Note that the diagonal element $D_{m,m}(t)$ for a single particle problem gives the density profile $P(m,t)$ at time $t$. We write down the equation of motion for $D_{m,n}(t)$ following the LQME in Eq.~\eqref{QME} and obtain 
\begin{eqnarray}
    \frac{d}{dt}D_{m,n}(t)&=&i J\sum_{l \ne0}\left(\frac{D_{m,n+l}(t)-D_{m+l,n}(t)}{|l|^\alpha}\right) \nonumber \\
    &+&\kappa\,(\delta_{m,n}-1)D_{m,n}(t).    
\end{eqnarray}
In the strong dephasing limit $(\kappa \gg J)$, and using a perturbative expansion of $D_{m,n}$ in terms of a small parameter $\epsilon$, we obtain the following fractional-diffusion equation for the density profile (see Appendix \ref{app:fde} for the details of the derivation) 
\begin{equation}
    \frac{d}{dt}D_{m,m}(t)=\frac{2 J^2}{\kappa}\sum_{l \ne0}\left(\frac{D_{m+l,m+l}(t)-D_{m,m}(t)}{|l|^{2\alpha}}\right).
    \label{frac-diff}
\end{equation}
Note that, in the nearest neighbour case ($\alpha  \to \infty$), the summation in Eq.~\eqref{frac-diff} is restricted to $l= \pm 1$ which leads to a conventional diffusion equation with diffusion constant $\Delta=2 J^2/\kappa$. Remarkably, Eq.~\eqref{frac-diff} possesses interesting scaling forms \cite{schuckert2020nonlocal} in both the regimes $1 < \alpha < 1.5$ and $\alpha>1.5$. The scaling functions are given by:
\begin{eqnarray}
f\bigg(z=\frac{x}{(D_\alpha t)^{\eta}}\bigg)\approx \begin{cases}
F_\alpha(z),\hspace{1.2cm} 1<\alpha<1.5 \\
G(z),\hspace{1.4cm}\alpha>1.5
\label{scaling-form}
\end{cases}
\end{eqnarray}
with  $\eta = 1/(2 \alpha -1)$ for $1 < \alpha < 1.5$, and  $\eta = 1/2$ for  $\alpha>1.5$. Here, 
\begin{eqnarray}
\label{eq:FG}
F_\alpha(z)&=&\int_{-\infty}^{\infty} \frac{dk}{2\pi}e^{-|k|^{2 \alpha-1}}e^{izk},\quad \,
G(z)=\frac{e^{-z^2/4}}{2\sqrt{\pi}}. \,\,\,\,
\end{eqnarray}
There are finite time corrections to these scaling functions \cite{schuckert2020nonlocal}, which lead to heavy tails even when $\alpha>1.5$, i.e., corrections in $G(z)$ given in Eq.~\eqref{eq:FG} (see Appendix \ref{app:B} for details). $D_\alpha$ is a generalized diffusion constant, which is given by $D_\alpha\approx-2\Delta\Gamma(1-2\alpha)\sin(\pi\alpha)$ for $1<\alpha<1.5$ with $\Gamma$ being the Gamma function, while $D_\alpha\approx\Delta/(2\alpha-3)$ for $\alpha>1.5$. For $\alpha \gg 1.5$, the heavy tails would vanish and $D_\alpha\approx\Delta$. In Fig.~\ref{fig:collapse1pt2}, we demonstrate remarkable agreement between the scaling form (with no phenomenological parameters) given in Eq.~\eqref{scaling-form} (with corrections) and our extensive numerical simulations based on stochastic unraveling Eq.\eqref{noise-H}
. 

An important question that naturally emerges is the relation between the system-size scaling exponent of steady state conductance ($\delta$) and the exponent $\eta$ associated with the space-time collapse. We note that the anomalous super-diffusive transport in long-range systems is often governed by L\'evy flights \cite{schuckert2020nonlocal,Abhishek_Levy, Iubini_2022}. In our case, we find an intriguing connection with a well known random walk model in low-dimensional systems, i.e., L\'evy walk \cite{RevModPhys.87.483}. The typical space-time scaling in the central region of a pulse dictated by L\'evy walker is $x \sim t^{1/\beta}$ where $\beta$ is the exponent associated with the time of flight distribution of a L\'evy walker \cite{Abhishek_Levy}. If such a system is connected to boundary reservoirs, then the system size scaling of conductance \cite{Abhishek_Levy} is given by $N^{1-\beta}$. For our setup, following the time dynamics of single particle density profile, we find that $\beta=2\alpha-1$ [Eq.~\eqref{scaling-form}], which immediately gives a relation between the exponents $\delta$ and $\alpha$ as $\delta= 2 \alpha -2 $. This perfectly matches with our numerical predictions on transport.  

{\it Summary.}
In summary, we have studied quantum transport properties in one-dimensional long-range fermionic system subjected to dephasing noise. The interesting interplay between the incoherent dephasing mechanism and the coherent long-range hopping, results in an anomalous behaviour. This clear departure from conventional diffusive transport is manifested both in NESS transport and in density profile dynamics. An interesting byproduct of our finding is that when $\alpha=5/4$, we get $\eta=2/3$ which is associated with the Kardar-Parisi-Zhang universality class \cite{PhysRevLett.56.889,TAKEUCHI201877,spohn2014nonlinear,prahofer2004exact} at least as far as exponents in space-time correlations are concerned. Furthermore, the density profile dynamics was shown to emerge from a fractional diffusion equation which was derived following the multiple scale analysis technique. This aided in further cementing the relationship between conductance scaling exponent $\delta$ and the long-range hopping exponent $\alpha$.

Recent theoretical \cite{schuckert2020nonlocal} and experimental \cite{joshi2022observing} advances in interacting quantum spin-chains have reported such anomalous transport by studying unequal space-time spin-spin correlators. Our work reveals a possible intriguing connection between such interacting quantum systems and systems subjected to dephasing noise mechanisms. Bringing out this interesting connection is an important future direction. The complex interplay between long-range hopping, dephasing noise in the bulk and boundary reservoirs can have interesting implications in several quantities such as entanglement entropy \cite{kehrein2024page,saha2024generalised,PhysRevD.109.L081901}, negativity \cite{plenio2005logarithmic}, and quantum fluctuations \cite{PhysRevB.83.161408,PhysRevLett.102.100502}.

\section*{Acknowledgements}
A.D and B. K. A. would like to thank Devendra Singh Bhakuni for insightful discussions. K. G., M. K.  and B. K. A would like to thank Madhumita Saha for extensive discussions on the project. M. K. thanks Vishal Vasan for useful discussions. B. K. A. would like to acknowledge funding from the National Mission on Interdisciplinary  Cyber-Physical  Systems (NM-ICPS)  of the Department of Science and Technology,  Govt.~of  India through the I-HUB  Quantum  Technology  Foundation, Pune, India. K.G. would like to acknowledge the Prime Minister's Research Fellowship (ID- 0703043), Government of India for funding. 
M.K. would like to acknowledge support from the project 6004-1 of the Indo-French Centre for the Promotion of Advanced Research (IFCPAR).  M. K. thanks the VAJRA faculty scheme (No.~VJR/2019/000079) from the Science and Engineering Research Board (SERB), Department of Science and Technology, Government of India. M. K. acknowledges support of the Department of Atomic Energy, Government of India, under Project No. RTI4001. M. K. thanks the hospitality of the Department of Physics, University of Crete (UOC) and the Institute of Electronic Structure and Laser (IESL) - FORTH, at Heraklion, Greece.

\bibliography{References}

\newpage
\appendix
\onecolumngrid

\section{Derivation of the fractional diffusion equation from the Lindblad QME in Eq.~\eqref{QME}}
\label{app:fde}
In this appendix, we provide a derivation for the fractional diffusion equation for the long-range setup following the Lindblad QME given in Eq.~\eqref{QME}. The derivation is based on the multiple scale analysis technique \cite{Bender1999}. An alternative route is that of renormalization group technique which was recently employed to derive a diffusion equation \cite{universal_dissipation}. We start with the time evolution of the density matrix which is given by,
\begin{equation}
    \frac{d\hat{\rho}(t)}{dt}=-i[\hat{\cal H}_{S},\hat{\rho}(t)]+\kappa \sum_{i=(-N +1)/2}^{(N-1)/2} \left(\hat{n}_i\rho(t) \hat{n}_i-\frac{1}{2}\{\hat{n}_i^2,\hat{\rho}(t)\}\right).
    \label{LQME}
\end{equation}
Here we have consider the middle site of the lattice as origin. $\hat{\cal H}_S$ is the long-range lattice Hamiltonian as given in Eq.~\eqref{long-range-Ham}, $\kappa$ is the effective dephasing strength. In what follows, we show that in the large dephasing $\kappa$ limit, the single-particle density profile obeys a fractional diffusion equation. For that purpose, we focus on the correlator $D_{m,n}(t)\equiv \langle \hat{c}_m^\dagger(t)\hat{c}_n(t) \rangle$ and write down its equation of motion following the LQME in Eq.~\eqref{LQME}. Note that, for the single particle problem the diagonal elements of $D_{m,m}(t)$ gives the density profile $P(m,t)$ at time $t$. We obtain,
\begin{align}
\begin{split}
    \frac{d}{dt}D_{m,n}(t)=&i J \sum_{l\ne0}\left(\frac{D_{m,n+l}(t)-D_{m+l,n}(t)}{|l|^\alpha}\right)
    +\kappa\,(\delta_{m,n}-1)D_{m,n}(t).
\end{split}
\end{align}
We change the variables to $\tau=\kappa \,t$, and work in the strong dephasing limit i.e., $\kappa \gg J$ where recall that $J$ is the hopping strength. We introduce a parameter $\epsilon_l=\frac{J\kappa^{-1}}{l^\alpha}$, with $l\ge1$ and receive,
\begin{align}
    \frac{d}{d\tau}D_{m,n}(\tau)=&i\sum_{l\ne0}\epsilon_{|l|}\, \bigg(D_{m,n+l}(\tau)-D_{m+l,n}(\tau)\bigg)
    +\lambda_{m,n} D_{m,n}(\tau),
    \label{dmntau}
\end{align}
where $\lambda_{m,n}=\delta_{m,n}-1$. We will expand the solution $D_{m,n}$ in terms of a small parameter
\begin{equation}
    \epsilon=\sum_{l\ge1}\epsilon_l= \frac{J}{\kappa} \xi(\alpha),
\end{equation}
where $\xi(\alpha)$ is the Riemann-Zeta function which converges for $\alpha>1$. We first seek for a convergent solution for $D_{m,n}(\tau)$ by expanding it in powers of $\epsilon$ as, 
\begin{equation}
    D_{m,n}(\tau)=D_{m,n}^{(0)}(\tau)+\epsilon \, D_{m,n}^{(1)}(\tau)+\epsilon^2\, D_{m,n}^{(2)}(\tau)+...
    \label{perturbative}
\end{equation}
Substituting this in Eq.~\eqref{dmntau}, and matching order by order of $\epsilon$, we obtain:
\begin{align}
    \frac{d}{d\tau}D_{m,n}^{(0)}(\tau)=&\lambda_{m,n}D_{m,n}^{(0)}(\tau), \\
    \begin{split}
        \epsilon\frac{d}{d\tau}D_{m,n}^{(1)}(\tau)=&\epsilon\,\lambda_{m,n}\,D_{m,n}^{(1)}(\tau)
        +i\sum_{l\ne0}\epsilon_{|l|} \, \Big(D_{m,n+l}^{(0)}(\tau)-D_{m+l,n}^{(0)}(\tau)\Big),
    \end{split}
  \\
  \begin{split}
      \epsilon^2\frac{d}{d\tau}D_{m,n}^{(2)}(\tau)=&\epsilon^2\lambda_{m,n}D_{m,n}^{(2)}(\tau)
      +i\sum_{l\ne0}\epsilon_{|l|} \epsilon\left(D_{m,n+k}^{(1)}(\tau)-D_{m+k,n}^{(1)}(\tau)\right),
  \end{split}
 \\
    \vdots \\
    \begin{split}
        \epsilon^a\frac{d}{d\tau}D_{m,n}^{(a)}(\tau)=&\epsilon^a\lambda_{m,n}D_{m,n}^{(a)}(\tau)
        +i\sum_{l \ne0}\epsilon_{|l|} \epsilon^{a-1}\left(D_{m,n+k}^{(a-1)}(\tau)-D_{m+k,n}^{(a-1)}(\tau)\right),
    \end{split}
\end{align}
where the symbol $a>1$ indicates $a$-th order in $\epsilon$. We re-write the above equations as, 
\begin{align}
    \frac{d}{d\tau}D_{m,n}^{(0)}(\tau)=&\lambda_{m,n}D_{m,n}^{(0)}(\tau),\\
    \begin{split}
    \frac{d}{d\tau}D_{m,n}^{(1)}(\tau)=&\lambda_{m,n}D_{m,n}^{(1)}(\tau)
    +\frac{i}{\epsilon}\sum_{l \ne0}\epsilon_{|l|} \,\left(D_{m,n+l}^{(0)}(\tau)-D_{m+l,n}^{(0)}(\tau)\right),
    \end{split}
    \\
    \begin{split}
    \frac{d}{d\tau}D_{m,n}^{(2)}(\tau)=&\lambda_{m,n}D_{m,n}^{(2)}(\tau)
    +\frac{i}{\epsilon}\sum_{l\ne0}\epsilon_{|l|} \,\left(D_{m,n+l}^{(1)}(\tau)-D_{m+l,n}^{(1)}(\tau)\right),
    \end{split}
    \\
    \vdots
    \\
    \begin{split}
    \frac{d}{d\tau}D_{m,n}^{(a)}(\tau)=&\lambda_{m,n}D_{m,n}^{(a)}(\tau)
        +\frac{i}{\epsilon}\sum_{l\ne0}\epsilon_{|l|} \left(D_{m,n+l}^{(a-1)}(\tau)-D_{m+l,n}^{(a-1)}(\tau)\right).
    \end{split}
\end{align}
Let us now consider the initial condition to be such that $D_{m,n}(\tau=0)=D_{m,n}^{(0)}(\tau=0)$. This makes $D_{m,n}^{(a)}(\tau=0)=0$ for all $a>0$. Solving the $0^{th}$ order equation, we have
\begin{align}
\label{0th}
    D_{m,n}^{(0)}(\tau)=&A_{m,n}e^{\lambda_{m,n}\tau}, \\
    D_{m,n}^{(1)}(\tau)=&\frac{ie^{\lambda_{m,n}\tau}}{\epsilon}\int_0^\tau dt' \,e^{-\lambda_{m,n}t'}\sum_{l \ne0}\epsilon_{|l|} \left(D_{m,n+l}^{(0)}(t')-D_{m+l,n}^{(0)}(t')\right).
\end{align}
Now, substituting the $0^{th}$ order solution into $1^{st}$ order equation, we get for the diagonal and non-diagonal elements of $D^{(1)}_{m,n}(\tau)$ as 
\begin{align}
\label{1st-first}
    D_{m,m}^{(1)}(\tau)=&\frac{i(1-e^{-\tau})}{\epsilon}\sum_{l\ne0}\epsilon_{|l|} \left(A_{m,m+l}-A_{m+l,m}\right) \\
    \begin{split}
        D_{m,m+b}^{(1)}(\tau)=&\frac{i\tau e^{-\tau}}{\epsilon}\sum_{l\ne0,-b}\epsilon_{|l|}\left(A_{m,m+b+l}-A_{m-l,m+b}\right)
        +\frac{i(1-e^{-\tau})\,\epsilon_{|b|}}{\epsilon}\left(A_{m,m}-A_{m+b,m+b}\right)
    \end{split}
    \label{1st-second}
\end{align}
Now, because we are interested in the populations (upto 2nd order), we focus on $D_{m,m}^{(2)}(\tau)$
\begin{align}
    D_{m,m}^{(2)}(\tau)=&\frac{i}{\epsilon}\int_0^\tau dt'\,\sum_{l\ne0}\epsilon_{|l|} \, \left(D_{m,m+l}^{(1)}(t')-D_{m+l,m}^{(1)}(t')\right) \nonumber \\
    \begin{split}
    =&\frac{2\,\tau}{\epsilon^2}\sum_{l\ne0}\epsilon_{|l|}^2\left(A_{m+l,m+l}-A_{m,m}\right) + \text{(Convergent term in $\tau$)}
    \label{div_sol}
\end{split}
\end{align}
From Eq.~\eqref{div_sol}, it is clear that the second order term $D^{(2)}_{m,m}(\tau)$ diverges linearly with $\tau$. To circumvent this issue, we now adapt the multiple scale analysis technique.

Let us seek for a solution 
\begin{equation}
    D_{m,n}(\tau)=D_{m,n}^{(0)}(\tau,\tau')+\epsilon \, D_{m,n}^{(1)}(\tau,\tau')+\epsilon^2\, D_{m,n}^{(2)}(\tau,\tau')+...
\end{equation}
where we define a new independent time scale $\tau'= \epsilon^2 \tau$. Note that here we consider $D^{(0)}, D^{(1)}, D^{(2)}$ as functions of $\tau$ and $\tau'$. Since $d\tau'/d\tau = \epsilon^2$, we receive,
\begin{equation}
\frac{d}{d\tau} D_{m,n}(\tau) = \frac{\partial}{\partial \tau} D_{m,n}^{(0)}(\tau,\tau') + \epsilon \frac{\partial}{\partial \tau} D_{m,n}^{(1)} 
(\tau,\tau')+ \epsilon^2 \Big[ \frac{\partial}{\partial \tau} D_{m,n}^{(2)}(\tau,\tau') +  \frac{\partial}{\partial \tau'} D_{m,n}^{(0)}
(\tau,\tau')\Big]
\end{equation}
Now using Eq.~\eqref{dmntau} and matching terms order by order we get the following. For $0$-th order 
\begin{equation}
\frac{\partial}{\partial \tau}D_{m,n}^{(0)}(\tau,\tau')=\lambda_{m,n}D_{m,n}^{(0)}(\tau,\tau')
\label{0th}
\end{equation}
Integrating Eq.~\eqref{0th}, the solution of which is given as
\begin{equation}
D_{m,n}^{(0)}(\tau,\tau') =A_{m,n}(\tau') \, e^{\lambda_{m,n} \tau}
\label{Dmn0}
\end{equation}
where recall that $\tau'= \epsilon^2 \tau$. In order $\epsilon$, we receive, 
\begin{equation}
\begin{split}
        \frac{\partial}{\partial \tau}D_{m,n}^{(1)}(\tau,\tau')=& \lambda_{m,n}\,D_{m,n}^{(1)}(\tau,\tau')
        +\frac{i}{\epsilon} \sum_{l\ne0}\epsilon_{|l|} \, \Big(D_{m,n+l}^{(0)}(\tau,\tau')-D_{m+l,n}^{(0)}(\tau,\tau')\Big),
    \end{split}
\end{equation}
which can be solved as before and the solution for $D_{m,n}^{(1)}(\tau,\tau')$ is given by Eq.~\eqref{1st-first} and Eq.~\eqref{1st-second} with $A$ now having a dependence on $\tau'$. Now for the second order in $\epsilon$, we receive,
\begin{equation}
\frac{\partial}{\partial \tau}D_{m,n}^{(2)}(\tau,\tau')= \lambda_{m,n}\,D_{m,n}^{(2)}(\tau,\tau') + \frac{i}{\epsilon} \sum_{l\ne0}\epsilon_{|l|} \, \Big(D_{m,n+l}^{(1)}(\tau,\tau')-D_{m+l,n}^{(1)}(\tau,\tau')\Big) - \frac{\partial}{\partial \tau'}D_{m,n}^{(0)}(\tau,\tau').
\label{Dmn2}
\end{equation}
Let us investigate the diagonal elements of $D_{m,m}^{(2)}(\tau,\tau')$ in Eq.~\eqref{Dmn2}, which gives, 
\begin{equation}
\frac{\partial}{\partial \tau}D_{m,m}^{(2)}(\tau,\tau')= \frac{i}{\epsilon} \sum_{l\ne0}\epsilon_{|l|} \, \Big(D_{m,m+l}^{(1)}(\tau,\tau')-D_{m+l,m}^{(1)}(\tau,\tau')\Big) - \frac{d}{d \tau'}A_{m,m}(\tau'),
\end{equation}
where we have used the fact that $D_{m,m}^{(0)}(\tau,\tau') =A_{m,m}(\tau')$ from Eq.~\eqref{Dmn0}. Upon substituting the solution for $D_{m,m+l}^{(1)}(\tau,\tau')$ and $D_{m+l,m}^{(1)}(\tau,\tau')$ following Eq.~\eqref{1st-second}, we receive, 
\begin{equation}
\frac{\partial}{\partial \tau}D_{m,m}^{(2)}(\tau,\tau')=  \Bigg[\frac{2}{\epsilon^2} \sum_{l\ne0}\epsilon_{|l|}^2\Big(A_{m+l,m+l}(\tau')-A_{m,m}(\tau') \Big) - \frac{d}{d \tau'}A_{m,m}(\tau') \Bigg] + \text{(Convergent term in $\tau$)}.
\label{Amm}
\end{equation}
Note that the term $\Big(A_{m+l,m+l}(\tau')-A_{m,m}(\tau') \Big)$ in Eq.~\eqref{Amm} would be the origin of divergent solution. This is because this term appears in Eq.~\eqref{1st-second} without the exponentially suppressed factor in $\tau$. Therefore, to ensure that the solution of Eq.~\eqref{Amm} is convergent, we impose,
\begin{equation}
 \frac{d}{d \tau'}A_{m,m}(\tau') = \frac{2}{\epsilon^2} \sum_{l\ne0}\epsilon_{|l|}^2\Big(A_{m+l,m+l}(\tau')-A_{m,m}(\tau') \Big).
\end{equation}
Following Eq.~\eqref{perturbative} and Eq.~\eqref{0th}, we immediately receive an equation for $D_{m,m}(\tau)$ given as 
\begin{equation}
    \frac{d}{d\tau}D_{m,m}(\tau)=2\sum_{l\ne0}\epsilon_{|l|}^2\bigg(D_{m+l,m+l}(\tau)-D_{m,m}(\tau)\bigg).
\end{equation}
Transforming back to the time variable $t$ as $\tau=\kappa t$, we get,
\begin{equation}
    \frac{d}{dt}D_{m,m}(t)=\frac{2J^2}{\kappa}\sum_{l\ne0}\left(\frac{D_{m+l,m+l}(t)-D_{m,m}(t)}{|l|^{2\alpha}}\right).
\end{equation}
This is the central equation which describes that a classical master equation for the population that satisfies a fractional diffusion equation.

\section{Finite time corrections to the scaling function}
\label{app:B}
In this appendix, we explicitly write down the scaling function for $\alpha>1.5$ with finite time corrections. The corrected scaling function $\Tilde{G}(z)$ is given by \cite{schuckert2020nonlocal}
\begin{equation}
 \Tilde{G}(z)=G(z)+(\sqrt{D_\alpha t})^{3-2\alpha}\frac{(2\alpha-3)\sin(\pi\alpha)\Gamma(1-2\alpha)\Gamma(\alpha)}{8\pi}{}_1F_1\big[\alpha,\frac{1}{2},-\frac{y^2}{4}\big].  
\end{equation}
Here, $z=x/\sqrt{D_\alpha t}$, $D_\alpha=\Delta/(2\alpha-3)$, and ${}_1F_1[\cdot,\cdot,\cdot]$ denotes Kummer confluent hypergeometric function, which has heavy tails $\sim z^{-2\alpha}$ for large $z$.
\end{document}